\documentclass[aip,pop]{revtex4-1} 
\usepackage{graphicx,bm}

\newcommand{\EQ}{\begin{equation}}
\newcommand{\EN}{\end{equation}}

\newcommand{\Fig}[1]{Figure~\ref{#1}}
\newcommand{\fig}[1]{figure~\ref{#1}}

\newcommand{\BB}{\mbox{\boldmath $B$} {}}
\newcommand{\bb}{\mbox{\boldmath $b$} {}}
\newcommand{\zz}{\mbox{\boldmath $z$} {}}
\newcommand{\JJ}{\mbox{\boldmath $J$} {}}
\newcommand{\nab}{\mbox{\boldmath $\nabla$} {}}
\newcommand{\vv}{\mbox{\boldmath $v$} {}}
\newcommand{\EE}{\mbox{\boldmath $E$} {}}

\newcommand{\PP}{\mbox{\boldmath $P$} {}}

\newcommand{\BBm}{\overline{\mbox{\boldmath $B$}}{}}{}
\newcommand{\BBf}{\tilde{\mbox{\boldmath $B$}}{}}{}
{}
\newcommand{\EEm}{\overline{\mbox{\boldmath $E$}}{}}{}
\newcommand{\EEf}{\tilde{\mbox{\boldmath $E$}}{}}{}
\newcommand{\vvm}{\overline{\mbox{\boldmath $v$}}{}}{}
\newcommand{\vvf}{\tilde{\mbox{\boldmath $v$}}{}}{}
\newcommand{\PPm}{\overline{\mbox{\boldmath $P$}}{}}{}
{}
\newcommand{\JJm}{\overline{\mbox{\boldmath $J$}}{}}{}
\newcommand{\JJf}{\tilde{\mbox{\boldmath $J$}}{}}{}
\newcommand{\nnm}{\overline{\mbox{$n$}}{}}{}
\newcommand{\nnf}{\tilde{\mbox{$n$}}{}}{}

\newcommand{\nat}{\em Nature \em}
\newcommand{\jfm}{\em J.\ Fluid Mech. \em}

\newcommand{\aap}{\em Astron.\ Astrophys. \em}

\newcommand{\mnras}{\em Mon.\ Not.\ Roy.\ Astron.\ Soc. \em}

\newcommand{\jgr}{\em J.\ Geophys.\ Res. \em}
\newcommand{\phof}{\em Phys.\ Fluids \em}
\newcommand{\pop}{\em Phys.\ Plasmas \em}
\newcommand{\apj}{\em Astrophys.\ J. \em}

\newcommand{\apjs}{\em Astrophys.\ J.\ Suppl. \em}
\newcommand{\apjl}{\em Astrophys.\ J.\ Lett. \em}

\newcommand{\pre}{\em Phys.\ Rev.\ E \em}
\newcommand{\prl}{\em Phys.\ Rev.\ Lett. \em}
\newcommand{\ssr}{\em Space\ Sci.\ Rev. \em}
\newcommand{\azh}{\em Astron.\ Zhurnal \em}
\newcommand{\jopp}{\em J.\ Plasma\ Phys.  \em}
\newcommand{\ppcf}{\em Plasma\ Phys.\ Contr.\ Fusion \em}
\newcommand{\pss}{\em Plan.\ Space\ Sci. \em}
\newcommand{\grl}{\em Geophys.\ Res.\ Lett. \em}

\begin{document}

\title{Kinetic plasma turbulence during the nonlinear stage of the Kelvin-Helmholtz instability}

\author{Koen Kemel$^1$, Pierre Henri$^2$, Giovanni Lapenta$^1$, Francesco Califano$^3$, Stefano Markidis$^4$}
\affiliation{
$^1$ Centre for mathematical Plasma Astrophysics, Department of Mathematics, KU Leuven, 
Celestijnenlaan 200B, 3001 Heverlee, Belgium\\
$^2$ LPC2E, CNRS, Univ. d'Orl\'{e}ans, OSUC, Orl\'eans, France\\
$^3$ Dip. Fisica, Universit\`{a} di Pisa; Largo Pontecorvo 3, 56127 Pisa, Italy\\
$^4$ HPCViz Department, KTH Royal Institute of Technology, Stockholm, Sweden
}


\begin{abstract}

Using a full kinetic, implicit particle-in-cell code, iPiC3D, 
we studied the properties of plasma kinetic turbulence, 
such as would be found at the interface between the solar wind and the Earth magnetosphere at low latitude during northwards periods.
In this case, in the presence of a magnetic field B oriented mostly perpendicular to the velocity shear, 
turbulence is fed by the disruption of a Kelvin-Helmholtz vortex chain via secondary instabilities, vortex pairing and non-linear interactions.

We found that the magnetic energy spectral cascade between ion and electron inertial scales, $d_i$ and $d_e$, 
is in agreement with satellite observations and other previous numerical simulations; 
however, in our case the spectrum ends with a peak beyond $d_e$ due to the occurrence of the lower hybrid drift instability.
The electric energy spectrum is influenced by effects of secondary instabilities: 
anomalous resistivity, fed by the development of the lower hybrid drift instability, steepens the spectral decay and, 
depending on the alignment or anti-alignment of B and the shear vorticity, 
peaks due to ion-Bernstein waves may dominate the spectrum around $d_i$.
These waves are generated by counter-streaming flow structures, 
through flux freezing also responsible for reconnection of the in-plane component of the magnetic field, 
which then generates electron pressure anisotropy and flattening of the field-aligned component of the electron distribution function.

\end{abstract}

\maketitle

\section{Introduction}

With a large separation between the scales of system dynamics and dissipation, 
astrophysical and space flows are often in a turbulent state,
at the same time, they consist dominantly of fully ionised plasmas. 
A major implication of the previous sentence is the multiscale aspect of our descriptions and calculations.

Plasma species are subject to a variety of physical processes, to which one can associate a typical length scale (and a correlated time scale): 
collisional length scales, the scales at which waves and particle motions decouple (inertial length) and magnetisation scales (Larmor radius). 
Depending on temperature, density and field strength, these plasma scales can be ordered in various ways.  
In a turbulent flow, characterised by the nonlinear transport of energy between scales, an injection scale and dissipative scales are also to be taken into account. 
While in a fluid description the dissipation scale is a single length set by collisions,
in weakly collisional plasmas several energy redistribution paths can coexist, 
and energy dissipation is the result of a complex interaction of micro-instabilities. 

In Astronomy, one would like to think in terms of the larger (observable) scales and events, 
far above the necessary resolution set by the plasma-kinetic equations - and often above the scales associated with fluid turbulence as well.
However, to correctly predict the transport coefficients used in such a macro desciption, 
one has to look at the actual interaction scales.
In this paper, we focus on the energy and momentum transfer between solar wind and the magnetosphere near the equator 
when the earth ans solar wind magnetic fields are approximately in the same direction \cite{Southwood74,Chen93,Hasegawa04,Pope09,Masters09,Sundberg10}.
This plasma shear flow configuration is unstable to the Kelvin-Helmholtz instability 
and has been extensively studies by means of fluid simulations\citep{Nagano79,Miura82,Nakamura10,Henri12}
However, the scales on which the instability operates, are below the fluid limit and sheath width, 
growth rates or transport coefficients are governed by plasma kinetic laws. 

Typically plasma kinetic simulations are limited to describing very small scale physical phenomena, 
as explicit calculation imposes severe constraints on the resolution, 
enforcing us to resolve the electron associated temporal and spatial scales.
Working with an implicit formulation of the equations, 
those resolution restrictions are loosened \cite{brackbill1982implicit},
allowing us to solve larger systems at an affordable computational cost.
The current work attempts to bridge the range from the largest electron governed scales, 
becoming available in the latest satellite data \cite{Alexandrova08,Safrankova13},
to scales where one could consider the plasma as a single turbulent fluid.

There are different ways to self-consistently drive turbulent flows 
that have been used in numerical simulations: 
using an external driver to inject energy in the system \cite{gressel2008}, 
starting from the decay of a large amplitude perturbation \cite{delsordo2010} or letting an instability evolve nonlinearly \cite{dalziel1999}. 
In this work, we use this last approach, namely the nonlinear stage of the magnetized Kelvin-Helmholtz instability. 
The initial shear flow is intended here as a source of free energy to drive a turbulent stage 
in order to study the properties of magnetized plasma turbulence.

Our objectives here are threefold: first we wish to observe the consequences of working with a reduced resolution, 
enabled by using an implicit scheme, by comparing with simulations fully resolving kinetic scales \cite{Karimabadi13}.
Second, we are aiming at a sufficient separation of scales to achieve a clear energy cascade scaling, 
comparable to observational data.
And finally, we want to assess the observed kinetic effects in terms of more macroscopic behaviour.

The paper is divided as follows. After a brief review of plasma turbulence studies in section~\ref{section:Review}, 
the model used in this study is described in section~\ref{section:Nsetup} 
and the outcome of this full kinetic simulation is presented in section~\ref{section:results}. 
Results are analysed in section~\ref{section:disc}.
Our findings are concisely summarised in section~\ref{section:conc}.

\section{Review of plasma turbulence studies} \label{section:Review}
Let us first have a brief look at the general nature of turbulence in magnetised plasmas
to see what kind of behaviour we may expect in our own simulations, starting from the outer scale.

While our simulation does not have any collisions, 
given a sufficient domain size, it should be possible to retrieve the Alfv\'enic cascade observed in 2D magneto-hydrodynamic (MHD) turbulence, 
as it does not rely on collisions and exists deep into the kinetic regime\cite{Alex2009}.

In MHD context,'Turbulence' typically refers to the strong, balanced turbulence cascade of magnetic and specific kinetic energy in an incompressible plasma. 
Strong turbulence implies dominance of the non-linear term over the dissipative term, 
as opposed to weak turbulence \cite{GaltierNazarenko2000,Galtier2006,Mininni07,PerezBoldyrev08,BoldyrevPerez09, Saur02}.
Incompressibility results in a straightforward coupling between the evolution of velocities and magnetic fields in the MHD and electron-MHD (EMHD) subspectra, 
allowing us to describe the problem in Els\"asser variables $\zz^\pm=\vv_{(e)}\pm \bb$, i.e. using linear combinations of the (electron) velocity and the magnetic field. 
Studies into the effects of compressibility \cite{ChoLaz02,ChoLaz03,Alex2009,Kowal10}, 
concluded that the slow compressible modes passively exchange energy with the incompressible Alfv\'enic modes, 
while the fast modes are energetically less relevant, 
leaving the spectral scaling exponent unchanged. 
Balanced turbulence is defined by equation of the Els\"asser energies $\int{(\zz^+)^2}=E^+=E^-=\bb^2+\vv^2\pm 2\vv\cdot \bb$, 
implying the absence of cross-helicity. 
According to general consensus \cite{Lithwick07, BeresLaz08, PodestaBhat10, PerezBoldyrev09} unbalanced turbulence does not change the scaling, only relative amplitude of the Els\"asser energies.

While 3D MHD Alfv\'enic turbulence is dividing the scientific world \cite{GS95,Haugen04,Mininni07,PodestaBhat10, BeresLaz10,PerezBoldyrev10,Boldyrev11,Alexandrova13}, 
the scaling power of -3/2 for its 2D counterpart is well established from theory \cite{Iroshnikov63,Kraichnan65}, 
in numerics\cite{Biskamp1989,Politano1989,Biskamp2001} and in experiments \cite{Sommeria1986}.

For magnetic and electron kinetic energy in the EMHD regime a -7/3 scaling (kinetic Alfv\'en turbulence) was predicted and found in EMHD simulations, for 2D \cite{Biskamp96} as well as 3D \cite{Biskamp99} and in observations \cite{Bale05}. However, more recent solar wind observations suggest a steeper -8/3 spectrum \cite{Alexandrova08,Safrankova13}, 
for this mismatch several possible explanations exist\cite{Howes11,BoldyrevPerez12}. Explicit PIC simulations \cite{Karimabadi13} suggest that the -8/3 spectrum also exists in 2D. 
Corresponding to these different explanations, an electric field energy scaling with exponent -1/3\cite{Howes11} or -2/3\cite{BoldyrevPerez12} is expected.
At scales below the electron inertial length, the magnetic field spectrum drops strongly due to Landau damping \cite{Howes11}.


\section{Numerical setup} \label{section:Nsetup}

To drive a fully developed turbulent state starting from a shear layer requires an evolution on a time scale of many hundred of ion gyro-periods, much larger than the typical time scales that can be covered by standard Particle-in-Cell (PIC) methods. To overcome this problem and cover such a large period of time, we make use of the fully kinetic, fully electromagnetic Particle-in-Cell code iPIC3D \cite{iPIC3D}, which implements the moment implicit method to suppress numerical instabilities when using large simulation time steps \cite{brackbill1982implicit}. In the following, all quantities are normalized to ion quantities: the ion gyro-frequency, $\omega_{c,i}$, the ion inertial length, $d_i=c/\omega_{p,i}$ or $l_{i,i}=c/\nu_{p,i}$, and the Alfv\'en velocity $V_A$. A reduced ion-to-electron mass ratio $m_i/m_e = 64$ is used for computational reasons. 

We consider a 2D $(x,y)$ physical space with 3D vector fields corresponding, in phase space, to a 2D-3V configuration. The size of the numerical box is $L_x \times L_y = 75 \times 200$ using $N_x \times N_y = 1152 \times 3072$ grid points, for a spatial resolution of $dx = dy = 0.065 \ d_i = 0.52 \ d_e$ (the electron inertial length). 
The initial sheared velocity field $\mathbf U = U_y(x) \ \mathbf e_y$ is characterized by a double shear layer where the velocity varies from $-A_{eq}$ to $+A_{eq}$. Such a double shear layer is used in order to impose periodic boundary conditions, thus avoiding spurious effects driven by the boundary conditions. The shear layers are equispaced and located at $y_{c,1} = 50$ and $y_{c,2} = 150$. The velocity profile reads: 
\[ 
U_x(y) = A_{eq} \ \left[ \tanh \left( \frac{y - y_{c,1}}{ L_{eq} } \right) - \tanh \left( \frac{y - y_{c,2}}{ L_{eq} } \right) -1 \right]  ; 
\] 
where the maximum velocity field strength is $A_{eq} = 0.5$ corresponding to a velocity jump $\Delta U = 1$. 
We take a shear scale length of a few ion skin depths (or ion inertial lengths), namely $L_{eq} = 3$. 
Outside from the shear layers where the system is homogeneous, the initial magnetic field is 
$\mathbf B_{eq} = B_x \ \mathbf e_x + B_z \ \mathbf e_z$, where $\| \mathbf B_{eq}\| = 1$ and $B_z = 10 \ B_x$; the initial ion and electron thermal velocities, $V_{th,i} = 0.5$ and $V_{th,e} = 1.79$ respectively, are isotropic; the initial density is uniform and equal to one. 
Quasineutrality $n_i = n_e = n$ is imposed everywhere at the beginning of the simulation. 
The plasma beta is $\beta \simeq 0.3$, so that the ion inertial length and gyroradius are roughly of same order. 

One of the main difficulties in the kinetic modeling of shear flows is the choice of the initial conditions since only a few kinetic equilibria are known for shear flow configurations \cite{CaiStoreyNeubert1990PhFlB}. On the other hand, when using a force balance MHD-like equilibrium as initial condition, the tensor pressure reacts very fast  at scales not far from the ion kinetic scale \cite{Henrietal2013POP} thus introducing strong fluctuations in the system. For these reasons, to mimic situations where the shear length is a few ion Larmor radii wide, such as the magnetosheath - magnetosphere boundary, we chose to implement an ``extended two-fluid equilibrium'', that retains first order corrections in terms of Finite Larmor Radii (FLR) effects \cite{Cerrietal2013POP}. Although it is not a proper kinetic equilibrium, this setup has been proven to be particularly efficient in preventing the generation of unwanted artifacts driven by the kinetic reaction to an initial fluid-like setup \cite{Henrietal2013POP,Cerrietal2013POP}. This consideration is particularly important in the range of parameters used in this study. 
The profiles of the thermal velocities are set according to the profiles of the diagonal elements of the pressure tensor in the ``Cerri equilibrium''\cite{Cerrietal2013POP} and 
the particles are loaded with a weight corresponding to the equilibrium density profile. 

In each cell, 100 particles are loaded for each species, for a total of about 1 billion particles. The simulation ran for a few million CPU hours on 8192 cores on FERMI at CINECA until the system reaches a time $t \omega_{ci} \simeq 1000$. 

The numerical box dimension has been chosen so that the fastest growing mode is m=2, resulting in the formation of two Kelvin-Helmholtz vortices in each shear layer at the beginning of the nonlinear phase. The vortices are eventually disrupted during the nonlinear evolution of the system, forming two turbulent layers. We hereafter concentrate on the properties of plasma turbulence at the turbulent stage of the full kinetic simulation. 

\begin{figure*}\begin{center}
\includegraphics[scale=0.5]{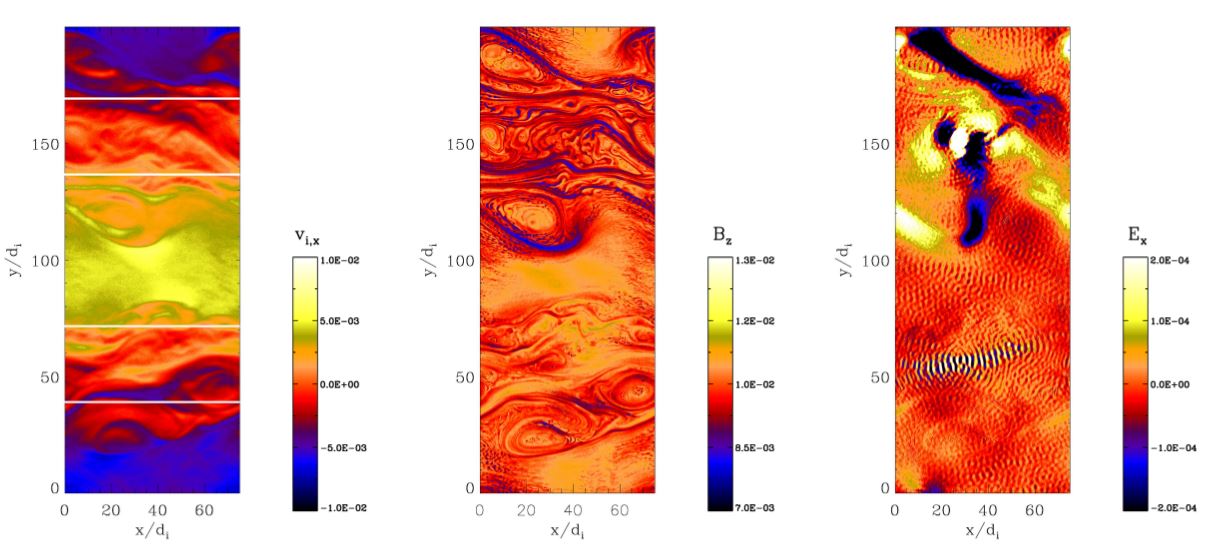}
\end{center}\caption[]{
Setup with 2 turbulent interaction layers, showing the horizontal component of the ion velocity (left), 
the out of the plane magnetic field (centre), and the horizontal component of the electric field (right).
White lines on the velocity plot indicate the definition of the turbulent layer location used in the spectral analysis.
}
\label{cplotsasym}\end{figure*}


\section{Results} \label{section:results}

\subsection{Effect of vorticity orientation} \label{AsymY}

One would expect a MHD setup with the same initial conditions for magnetic and flow fields to behave symmetric in the y-direction.
In \fig{cplotsasym} we see that especially for the electric field there are strong differences between the two turbulent layers, 
while for the ion velocity there is almost no difference.

\begin{figure*}\begin{center}
\includegraphics[scale=0.5]{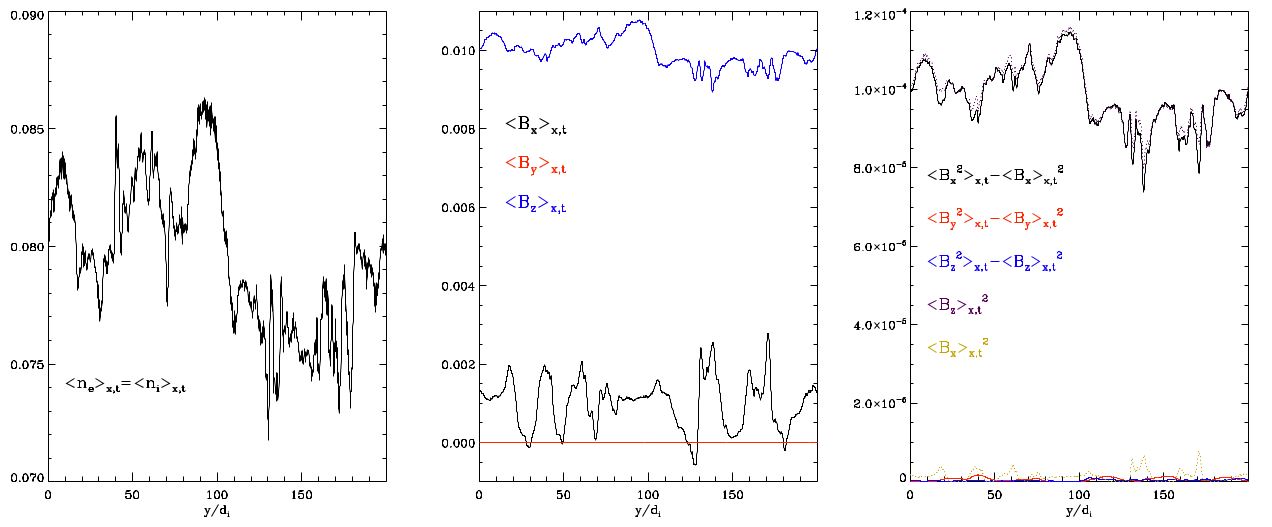}
\end{center}\caption[]{
Average over the horizontal direction and time of density (left), and the three components of the magnetic field (centre) 
and the energy in the fluctuations of B, i.e. $<B_j^2-<B_j>_{x,t}^2>_{x,t}$ (right).
}
\label{avBrho}\end{figure*}

\begin{figure*}\begin{center}
\includegraphics[scale=0.5]{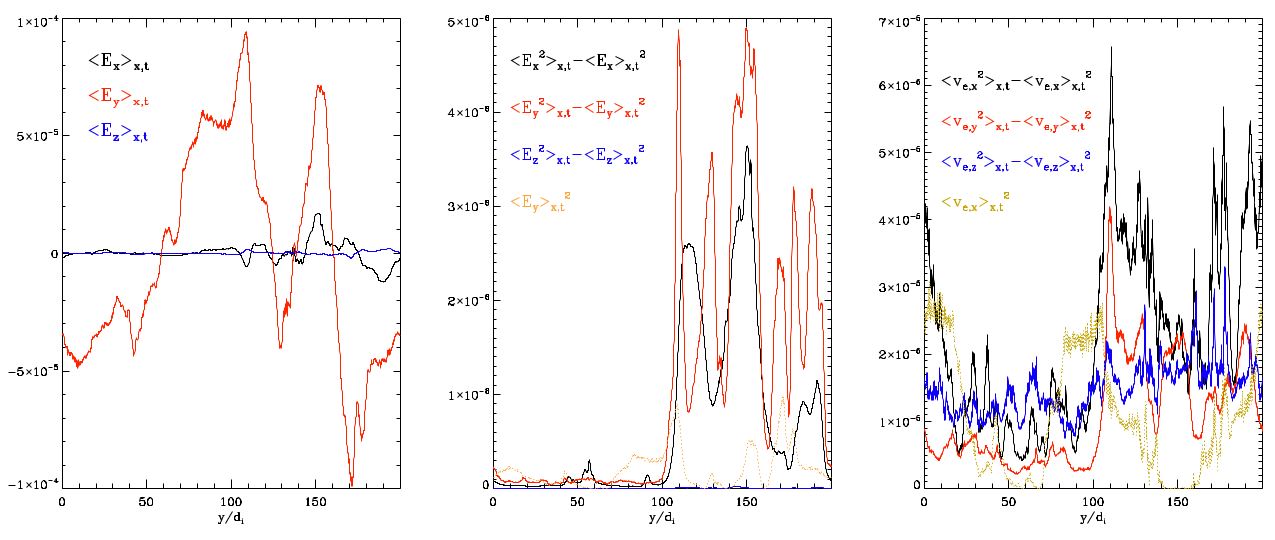}
\end{center}\caption[]{
Average over the horizontal direction and time of the three components of the electric field (left), 
the energy in the fluctuations of E, i.e. $<E_j^2-<E_j>_{x,t}^2>_{x,t}$ (centre), 
and in the fluctuations of the electron velocity, $<v_{e,j}^2-<v_{e,j}>_{x,t}^2>_{x,t}$ (right).
}
\label{avEE}\end{figure*}

The differences become more clear when we study the y-dependence by averaging over x and time. 
In \fig{avBrho} we see a mean density and magnetic field decrease in the upper part of the domain.
\Fig{avEE} shows a mean electric field corresponding to $v\times B$, modified in the top turbulent layer.
We find that the energy in the fluctuations (with a finite wavelength in x) 
in the top layer is much larger than the energy associated with the mean field,
while in the bottom layer the x-dependent contribution to the electric energy is weak. 

The vorticity of the shear flow in the top layer is in the same direction as the magnetic field.
This means that, for the electrons, the Lorentz force is pointing in the same direction as the acceleration by the flow, 
enhancing their rotational motions. The resulting electron currents also create a magnetic field that reduces the initially imposed one.
The magnetic field and flow are largely frozen-in, making the density follow the behaviour of B ($\beta<1$).
In the bottom layer, vorticity and B are anti-aligned and the forces on the electrons counteract, approximately cancelling out.
This results in asymmetry in the electron currents between the top and bottom of the domain, seen in \fig{avEE} on the right.
The electric energy shows similar asymmetric behaviour. 

\begin{figure*}\begin{center}
\includegraphics[scale=0.5]{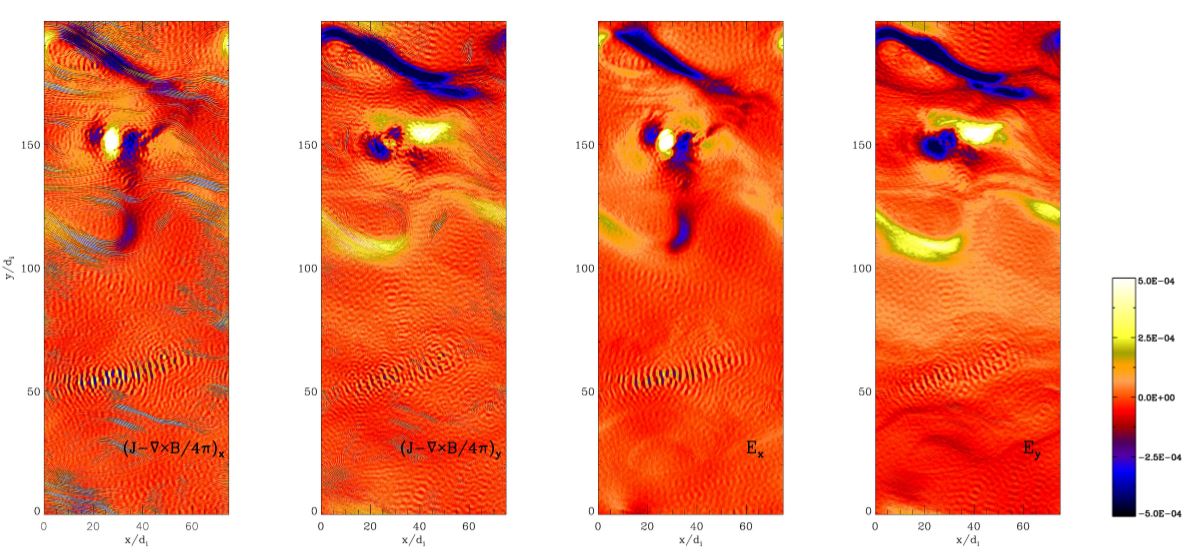}
\end{center}\caption[]{
Contour plot of the in-plane components of the displacement current (left) and of the electric field (right).
}
\label{cohm}\end{figure*}

To explain the similarity, this collisionless plasma would need some kind of resistivity.
A comparison of the spatial behaviour of electric field and current, \Fig{cohm}, does show a strong correlation.

Resistivity can be measured through taking an average of the (collisionless) momentum balance of the electrons:
$$m_e\frac{d\vv_e}{dt}=-(\EE+\vv_e\times\BB+\frac{1}{n_e}\nab\cdot\PP_e)$$
$$\nnm_e m_e\frac{d\vvm_e}{dt}+\nnm_e\EEm-\JJm_e\times\BBm+\nab\cdot\PPm_e=-(m_e<\nnf_e\frac{d\vvf_e}{dt}>+<\nnf_e\EEf>-<\JJf_e\times\BBf>)=\eta\JJm$$

\begin{figure*}\begin{center}
\includegraphics[scale=0.5]{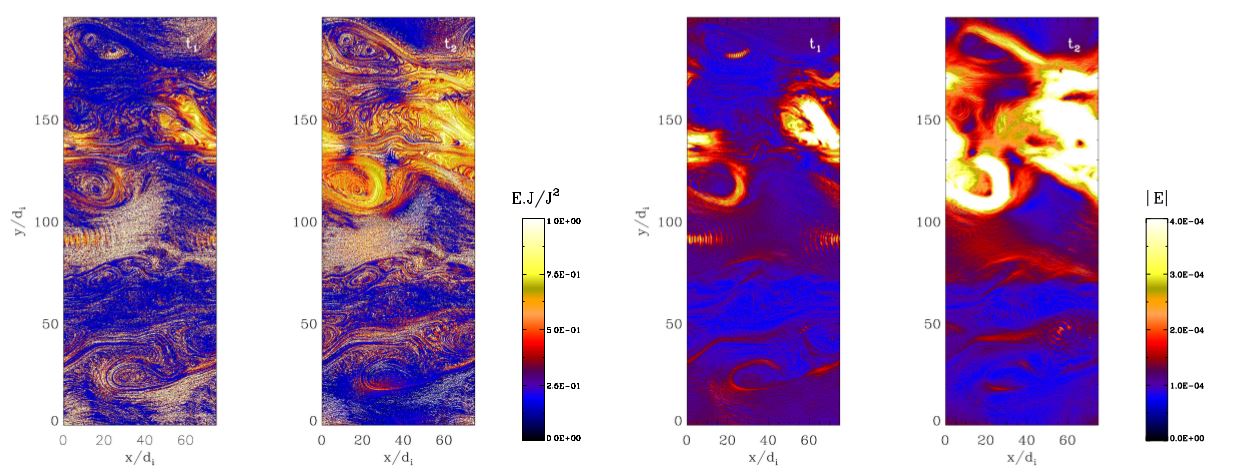}
\end{center}\caption[]{
Comparison of diffusivity estimate $\EE\cdot\JJ/J^2$ (left) with electric field structure $|E|$ (right)at two different times.
}
\label{fceta}
\end{figure*}

As the electron mass is small, we can drop the first term on the left, 
and the remaining terms are dominated by the electric field.
Estimating the diffusivity (actually a tensor) as the ratio $\EE\cdot\JJ/J^2$, 
we find, shown in \fig{fceta}, localised values around unity, 
surrounded by regions of weaker diffusivity (by approximately a factor 10).  
The most diffusive regions coincide with strong electric fields and small scale fluctuations.

\subsection{Spectral analysis}

We studied the 2D spectra of two subdomains within the upper and bottom turbulent layer. 
The non-periodic boundaries in y were mitigated by the use of a Hamming window, 
though for high angles with the x-axis some noise exists.

\begin{figure*}\begin{center}
\includegraphics[scale=0.5]{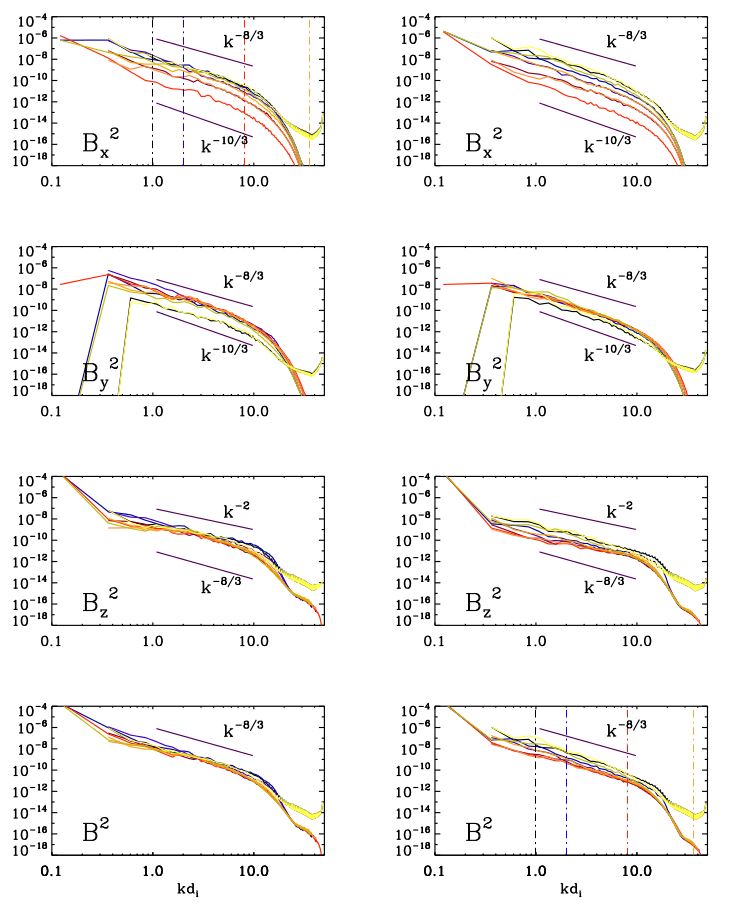}
\end{center}\caption[]{
Spectra for magnetic energy for different field components in the top (left) and bottom (right) turbulent layers.
Different colours indicate different angles in k-space, based on a $\pi/9$ bin around a central angle with the $k_x$-axis, 
ranging from $4\pi/9$ (black), over 0 (red) to $-4\pi/9$ (yellow).
Vertical lines indicate $kd_i=1$ (black), $k\rho_i=1$ (blue), $kd_e=1$ (red) and $k\rho_e=1$ (orange).
}\label{SpecB}\end{figure*}

In \fig{SpecB} we show the spectral scaling of the components of the magnetic field for different orientations.
The behaviour is largely isotropic and rather similar for both layers.
Scaling between the ion and electron scales is close to the expected value of -8/3, 
found also in Haynes et al. \cite{Haynes2014} and Karimabadi et al.\cite{Karimabadi13}.
Even though the domain size is on the order of $10^2d_i$, 
the resulting forcing scales provided by the instability are insufficiently large
to accommodate a fluid like energy cascade. 
One can see a slight bump in the spectrum of $B_z$, stronger for high angles, between $kd_e=1$ and $k\rho_e=1$.

\begin{figure*}\begin{center}
\includegraphics[scale=0.5]{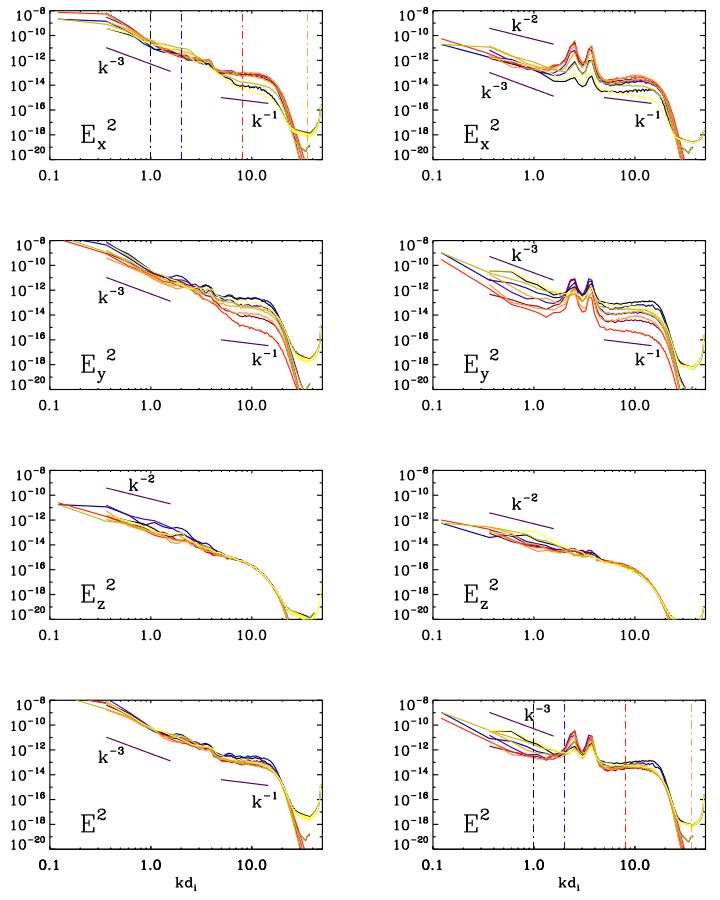}
\end{center}\caption[]{
Spectra for electric energy for different field components in the top (left) and bottom (right) turbulent layers.
Different colours indicate different angles in k-space, based on a $\pi/9$ bin around a central angle with the $k_x$-axis, 
ranging from $4\pi/9$ (black), over 0 (red) to $-4\pi/9$ (yellow).
Vertical lines indicate $kd_i=1$ (black), $k\rho_i=1$ (blue), $kd_e=1$ (red) and $k\rho_e=1$ (orange).
}\label{SpecE}\end{figure*}

\begin{figure*}\begin{center}
\includegraphics[scale=0.6]{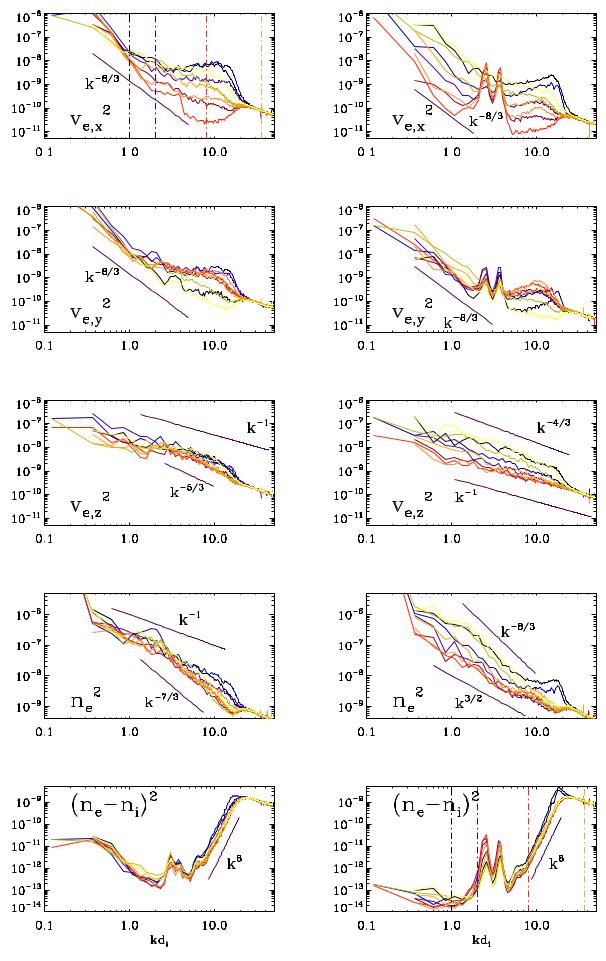}
\end{center}\caption[]{
Spectra for specific electron kinetic energy for different field components as well as electron density and its deviation from the ion density, in the top (left) and bottom (right) turbulent layers.
Different colours indicate different angles in k-space, based on a $\pi/9$ bin around a central angle with the $k_x$-axis, 
ranging from $4\pi/9$ (black), over 0 (red) to $-4\pi/9$ (yellow).
Vertical lines indicate $kd_i=1$ (black), $k\rho_i=1$ (blue), $kd_e=1$ (red) and $k\rho_e=1$ (orange).
}\label{SpecV}\end{figure*}

Contrary to the magnetic field spectrum, and as could already be anticipated from the previous section,
the k-space behaviour of the electric field is vastly different between the top and bottom layer
In the lower layer spectrum in \fig{SpecE}, we see two main features: a strong double peak, a bit beyond $k\rho_i=1$, and a bump, a bit beyond $kd_e=1$.
The bump is clearest in $E_y$ where, like in the magnetic spectrum, it is strongest for high angles. 
The double peak is strongest in $E_x$, where it favours small angles.
The scaling between the $d_i$ and $d_e$ is obscured by these two disturbances - scaling indications on the figures are for reference, rather than fits.
In the upper layer, traces of the double peak introduce noise on the measurement, though the, again largely isotropic, 
scaling is far steeper than expected for a collisionless plasma. 
The slope between -2 and -3 reminds of the magnetic spectrum and suggests the electric field follows from currents.   

The final set of spectra included here, \fig{SpecV} shows the electron velocity and density behaviour. 
The ion density matches the electrons closely, while all ion velocity components drop rather steeply around $kd_i=1$ down to noise levels, as in  Karimabadi et al.\cite{Karimabadi13}.
The velocity spectrum is largely dominated by the features already observed in the electric field: 
a double peak in $v_x$ along the $k_x$-axis a bit beyond $k\rho_i=1$ in the bottom layer (parallel to its strongest appearance in E) 
and a bump in $v_x$ 
(orthogonal to the orientation of its strongest appearances in both E and B), along the $k_y$-axis a bit beyond $kd_e=1$ in both layers.
The lowest four panels in \fig{SpecV} indicate the density scaling, we again notice an increase around the bump location in the other quantities. 
Only by looking at the scaling of the density difference between electrons and ions, do we recover the peaks at $kd_i=1$.

\subsection{Velocity distribution functions} \label{section:Vdist}

\begin{figure*}\begin{center}
\includegraphics[scale=0.5]{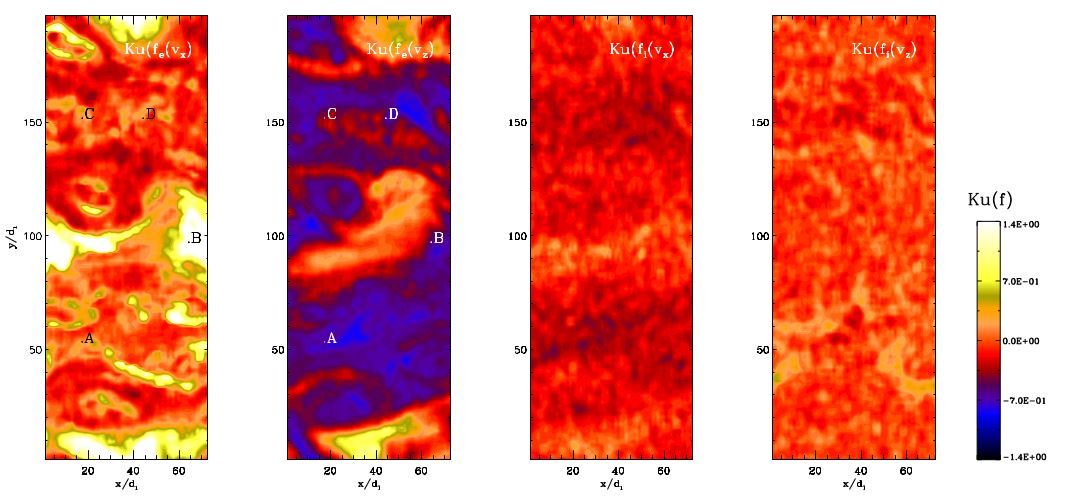}
\end{center}\caption[]{
Spatial dependence of the kurtosis of the distribution function, 
used as an indicator of non-maxwellian behaviour.
Left side: electrons, right side: ions; 
odd plots: based on dependence on $v_x$ , even plots: dependence on $v_z$ 
(A,B,C and D: see \fig{fdisf}). 
}
\label{fmoments}
\end{figure*}

\begin{figure*}\begin{center}
\includegraphics[scale=0.5]{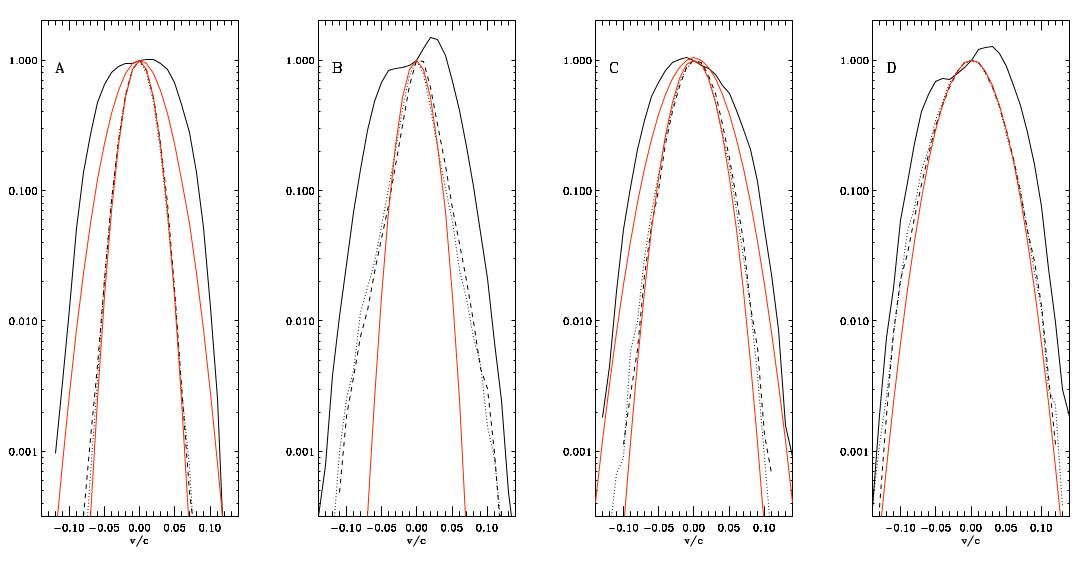}
\end{center}\caption[]{
Electron distribution functions corresponding to the four locations indicated in \fig{fmoments}, showing the dependence on $v_z$ (full line), $v_x$ (dashed) and $v_y$ (dotted) and Maxwellian approximations (red).
}
\label{fdisf}
\end{figure*}

\begin{figure*}\begin{center}
\includegraphics[scale=0.5]{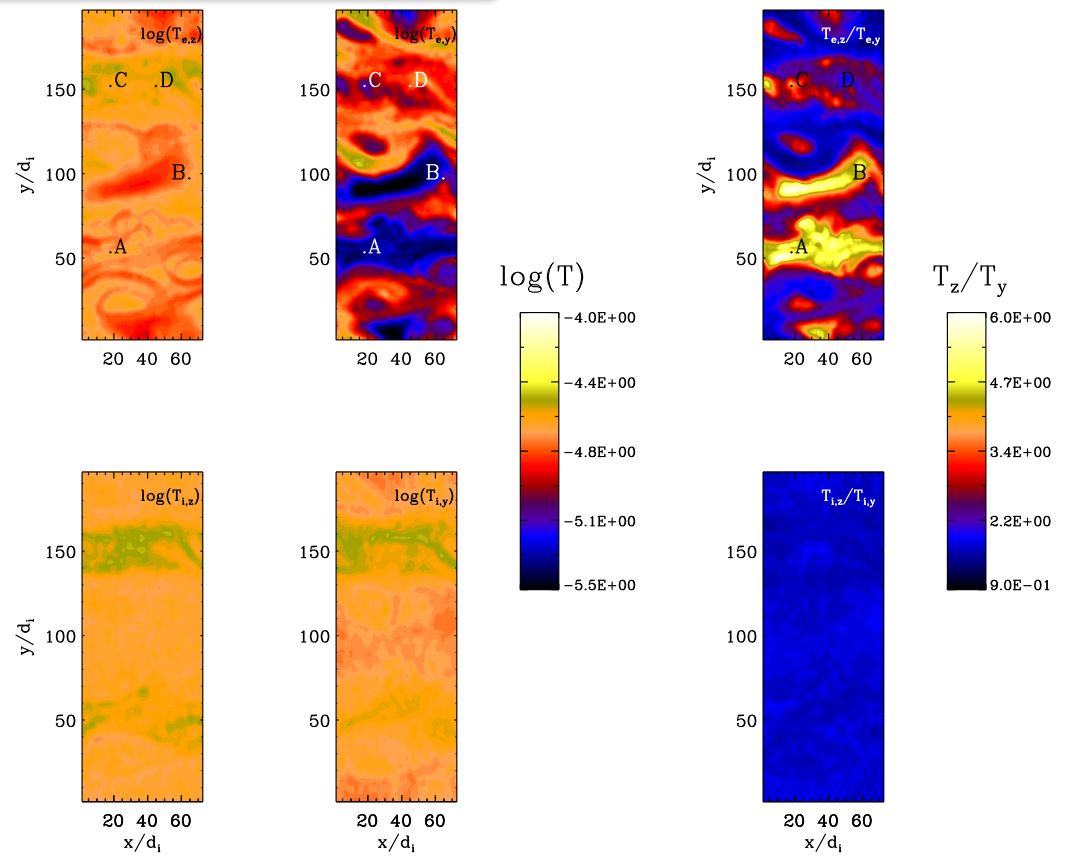}
\end{center}\caption[]{
Parallel (left) and perpendicular (centre) temperatures (variance of the distribution function along respectively $v_z$ and $v_y$) and their ratio (right), 
for electrons (top) and ions (bottom)(A,B,C and D: see \fig{fdisf}).
}
\label{temprat}
\end{figure*}

We calculated the velocity distribution function on a 64 by 128 grid, corresponding to the processor distribution.
In \fig{fmoments}, showing the spatial variation of the kurtosis for the distribution function along respectively $v_y$ and $v_z$, 
we see that for the electrons neither orientation is fully Maxwellian.
The distribution function is shown in \fig{fdisf} for different locations in the plane.
The velocity distribution in the plane is approximately isotropic and appears Maxwellian with a high energy tail of varying strength,  
while $f_e(v_z)$ is a flat-top distribution.
An approximation of the temperature ratio between the in and out of the plane dependances of the distribution function, using the corresponding variances (see 
\fig{temprat}) shows regions with large ratios of parallel to perpendicular temperature for the electrons.
The ion velocity distribution is approximately Maxwellian and does not show strong anisotropies. 

\section{Discussion} \label{section:disc}

\subsection{Pressure anisotropy and electrostatic waves}
\begin{figure*}\begin{center}
\includegraphics[scale=0.5]{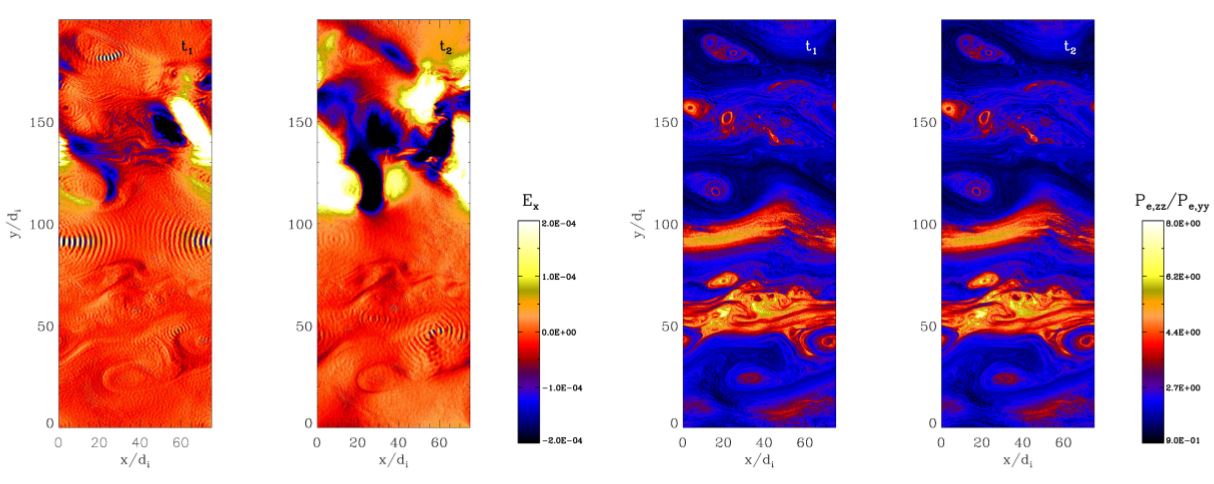}
\end{center}\caption[]{
Comparison of the electric field (left) and pressure anisotropy (right) at different times. 
Electrostatic waves are localised where pressure anisotropy is strong. 
Note that the fields change on a much shorter time scale than the pressure.
}
\label{cexpop}
\end{figure*}

Visually we observed, and through the spectra we confirmed the presence of electrostatic waves with preferred propagation along $x$.   
The waves are longitudinal and largely perpendicular to the magnetic field. 
They occur in regions with electron pressure anisotropy (the ion pressure is approximately isotropic), 
namely where the pressure component along B is larger, by a factor of about 5, than the pressure components perpendicular to B, see \fig{cexpop}.

The double peak in the electric energy spectrum, \fig{SpecE}, suggests the presence of the first and second proton cyclotron harmonic Bernstein waves. 
For propagation at high but less than perpendicular angle with the magnetic field these can result in a double peaked spectrum, 
becoming a single flat peak for orthogonal propagation \cite{LiHabbal2001, SahraouiBelmontGoldstein2012}. 
The spectral peaks for propagation along $y$ - generally at a higher angle with B are less pronounced and suggest a combination of the two spectra mentioned.

A cut through Kelvin-Helmholtz-rolls will show alternating flow directions. 
It was suggested \cite{ReynoldsGanguli1998} that such a geometry would become unstable to ion-Bernstein waves for layer separations on the order of the ion Larmor radius.

With approximately frozen-in magnetic fields, alternating flows correspond to alternating magnetic fields and hence reconnection. 
In earlier studies \cite{Egedal2002,EgedalLeDaughton2012} magnetic reconnection is linked to flattening of the electron distribution and electron pressure anisotropy of the same order as observed here. 

\begin{figure*}\begin{center}
\includegraphics[scale=0.5]{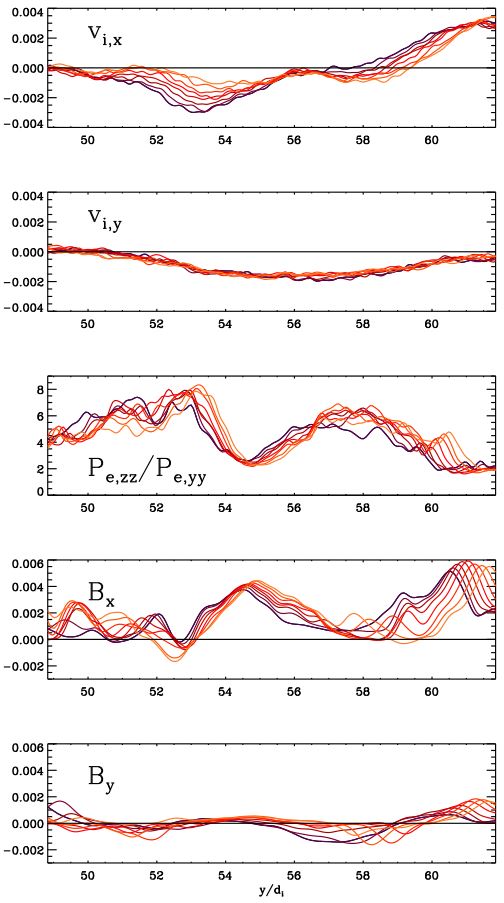}

\end{center}\caption[]{
Pressure anisotropy, in-plane velocity and magnetic field, along the $y$-direction 
going through region A indicated in \fig{temprat} and \fig{fdisf}, magnified in \fig{cz1epvb}, 
colours indicate different cuts slightly shifted along $x$. 
}
\label{fpvbp}
\end{figure*}

\begin{figure*}\begin{center}
\includegraphics[scale=0.5]{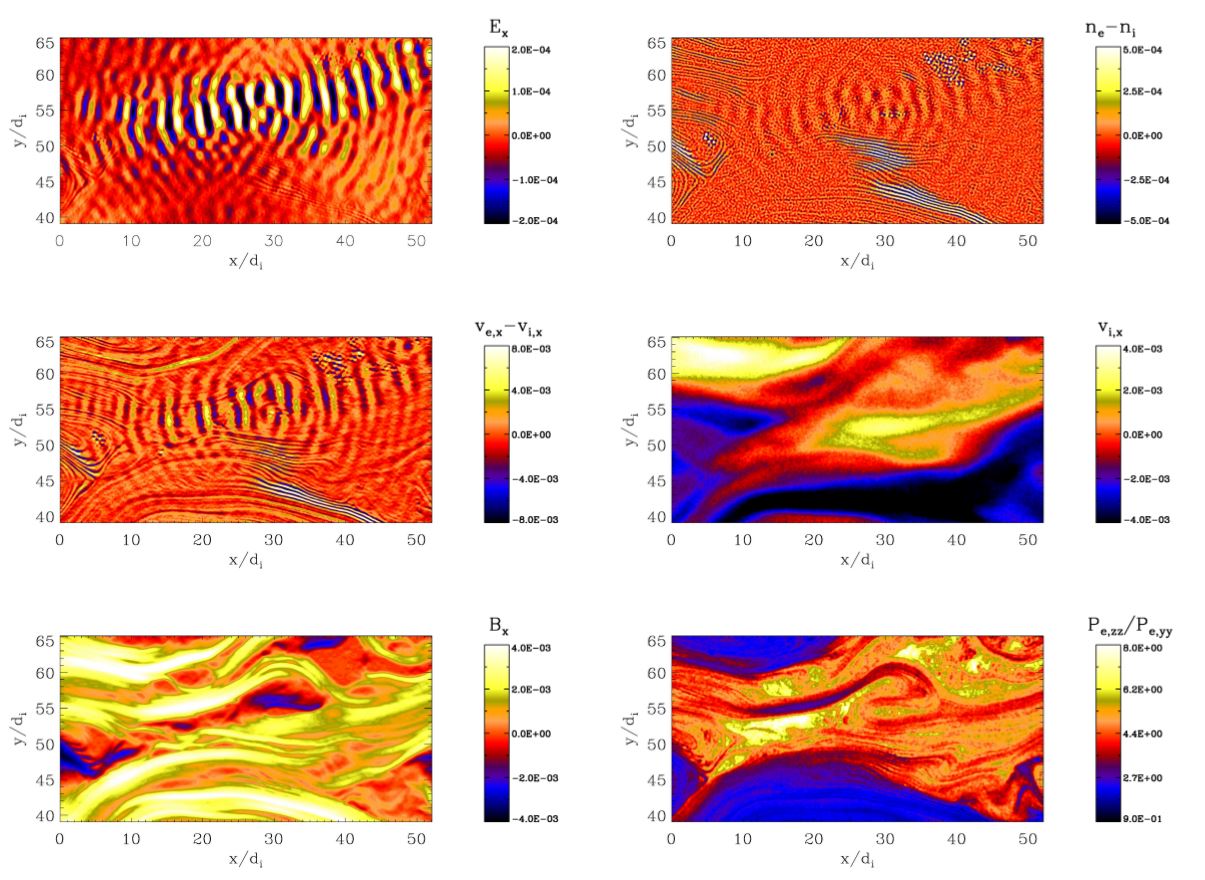}
\end{center}\caption[]{
Electric field (along $x$), density difference between electrons and ions,
relative electron velocity and ion velocity (along $x$), in the plane magnetic field (along $x$),
electron pressure anisotropy.
}
\label{cz1epvb}
\end{figure*}

While in the current stage of the instability evolution, there are no clean rolls, 
we do find similar velocity and field structures, see \fig{fpvbp}, in regions with strong anisotropy, 
ion-Bernstein-waves and a flat topped electron velocity distribution.
A contour plot for one of the regions of interest for the relevant quantities is given in \fig{cz1epvb}.

\subsection{Lower hybrid drift instability} 
\begin{figure*}\begin{center}
\includegraphics[scale=0.5]{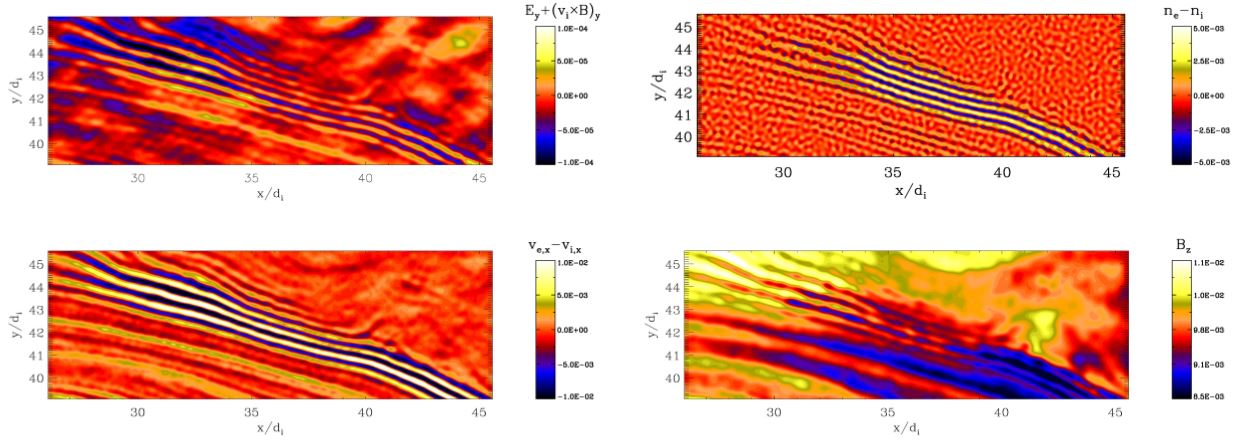}
\end{center}\caption[]{
Electric field (along $y$) with the effect of ion background motions removed, density difference between electrons and ions, 
relative electron velocity (along $x$) and out of the plane magnetic field.
}
\label{cz2envb}
\end{figure*}

In $E_y$,$B_z$,$v_{e,x}$ and $\rho$ we found the generation of waves with k-vector preferrably along y and norm between $kd_e=1$ and $k\rho_e=1$, shown in \fig{cz2envb}.
This signature agrees with the lower hybrid drift instability (LHDi).

The LHDi is typically associated with anomalous resistivity, which we already proposed in the asymmetry section, \fig{avEE}, and anticipated from the steep E-spectrum, \fig{SpecE}.
The LHDi identified regions indeed correspond to the stronger diffusivity locations in our estimate through Ohm's law, recall \fig{fceta}. 
The strongly localised diffusivity agrees with earlier simulations of the LHDi by Innocenti \& Lapenta (2007 \cite{innocenti2007}, and unpublished results).

\section{Conclusions} \label{section:conc}
While our domain size was of the order $10^2d_i$,
the turbulent injection scale, set by the Kelvin-Helmholtz instability, was
only a bit larger than $d_i$
and we were not able to resolve a fluid like spectral cascade.

For the magnetic field,
we were able to reproduce the kinetic Alfv\'en wave spectrum as seen in
existing simulations and observations.

The behaviour of the electric field is significantly different from what
one would expect from a cascade in a collisionless plasma.
The deviation is caused by two separate physical phenomena occuring on
respectively ion and electron scales.

In the region where the shear layer vorticity and magnetic field are
anti-aligned,
around $kd_i=2$,
an in-plane, alternating flow structure excites Ion-Bernstein waves,
resulting in a peaked electric energy spectrum.
This flow structure is also tied to the magnetic field structure, causing
reconnection of the in-plane magnetic field,
which, in turn, generates anisotropy in the electron velocity.

Troughout the whole domain,
near $kd_e=1$,
we find the lower hybrid drift instability (LHDI), showing up in both the
magnetic and the electric spectrum.
Beside the peak at the resonance wave length,
the LHDI also affects the electric field spectrum through the generation of
anomalous resistivity,
resulting in a far steeper slope.


\acknowledgments
\noindent The research leading to these results has received funding from the European Commission's Seventh Framework Programme (FP7/2007-2013) under the grant agreement SWIFF (project n 263340, www.swiff.eu).
We acknowledge the access to the HPC resources of CINECA made available within the Distributed European Computing Initiative by the PRACE-2IP, receiving funding from the European Community's Seventh Framework Programme (FP7/2007-2013) under grant agreement n$^\circ$ RI-283493 and project number 2012071282.
This work was supported by the Italian Super-computing Center CINECA under the ISCRA initiative.

\begin{scriptsize} 
\bibliographystyle{plainnat}
\renewcommand{\bibname}{References} 
\bibliography{references} 
\end{scriptsize}
\end{document}